\documentclass[aps,pra,twocolumn,floatfix,showpacs]{revtex4}
\usepackage{amsmath}
\usepackage{amsfonts}
\usepackage{amssymb}
\usepackage{graphicx}
\usepackage{color}

\begin{document}

\title{Strong squeezing via phonon mediated spontaneous generation of photon pairs}
\author{ Kenan Qu and G. S. Agarwal }
\affiliation{Department of Physics, Oklahoma State University, Stillwater, Oklahoma, 74078, USA}
\date{\today}

\begin{abstract}
We propose a scheme generating robust squeezed light by using double cavity optomechanical system driven by a blue detuned laser in one cavity and by a red detuned laser in the other. This double cavity system is shown to mimic effectively an interaction that is similar to the one for the downconverter, which is known to be a source of strong squeezing for the light fields. There are however distinctions as the phonons, which lead to such an interaction, can contribute to the quantum noise. We show that the squeezing of the output fields in the order of $10$dB can be achieved even at the mirror temperature at the order of $10$mK.
\end{abstract}
\maketitle

\section{Introduction}
Realization of the quantum regime~\cite{Girvin11,Marquardt,OMSEnt,double,Heidmann,Li,sqmirror,Stamper-Kurn,Gong,Girvin13,Hammerer,Clerk,Law,Asjad,Lehnert,Kirchmair,Ge} of physical systems has been of great interest. One important signature of the quantum regime is the squeezing which has been studied rather extensively for radiation fields~\cite{Wolf,Wallsbook,GSAbook}, collection of two level atoms and spins~\cite{spin1,spin2}, optical phonons~\cite{sood}, and Bose condensates~\cite{BEC0,BEC1,BEC2,BEC3}. The squeezing is of fundamental importance in precision measurements~\cite{CavesRMP,Braginsky,Yurke} and thus quantum metrology drives the demand not only for higher levels of squeezing but also for the availability of squeezing in a variety of systems. There are several studies demonstrating squeezing in optomechanical systems~\cite{sqmirror1,sqmirror2,sqmirror,squeezing,CavesPRL,Brooks,PRX,GSA}. In the present work on optomechanical systems, we develop an analog of the standard method for producing squeezing in quantum optics. We will also give a comparison of our proposal with other squeezing schemes.

In quantum optics, the downconverter is an important resource for producing squeezing~\cite{Wolf,Wallsbook,GSAbook}, thus it is desirable to have an optomechanical system which can behave effectively like a downconverter. In a downconverter, three light field modes defined by annihilation operators $a$, $b$ and $c$ get coupled in a nonlinear medium where the second order nonlinearity is nonzero. This coupling is described by the Hamiltonian $\chi a^\dag bc+\chi ab^\dag c^\dag$ which, under the condition of a strong undepleted pump field $a$, reduces to $\xi bc+\xi^* b^\dag c^\dag$. The photon pairs $b$ and $c$ are spontaneously produced from the pump. These are the entangled pairs. An appropriate linear combinations $(b+c\mathrm{e}^{i\theta})/\sqrt{2}$ of the modes $b$ and $c$ can produce quadrature squeezing. Note that the susceptibility $\chi$ for the downconverter has no resonances in the frequency range of interest. Any system whose effective interaction can be reduced to the form $\xi bc+\xi^* b^\dag c^\dag$ becomes a good candidate for squeezing. The third order nonlinearities in optical fibers can also give rise to such an interaction leading to a large number of studies on squeezing~\cite{Leuch}. In the cavity optomechanics, one can realize several parametric processes: A blue detuned pump with frequency $\omega_l$ can in fact produce spontaneously a cavity photon of frequency $\omega_c$ and a phonon of frequency $\omega_m$. In the undepleted pump approximation, this process would be described by an effective Hamiltonian $\xi bc+\xi^* b^\dag c^\dag$ where $b$ stands for the cavity photon and $c$ stands for the phonon. Though this Hamiltonian has the form of a downconverter, it cannot produce squeezing of the light field since $\omega_m$ is many orders less than $\omega_c$. Then one would like to replace the phonon mode by another optical mode. The cavity optomechanics has another parametric process where a red detuned pump and a probe, defined by annihilation operator $d$, can produce a coherent phonon. In the undepleted pump approximation, this is described by $\zeta d^\dag c+\zeta^* dc^\dag$. This interaction also implies that a phonon in combination with a red detuned pump will produce a cavity photon $d$. Thus a cavity photon can be produced using either the blue detuned pump or by using a red detuned pump and a phonon. In what follows, we use both these mechanisms to produce a pair of photons in a double cavity optomechanical system. Thus we produce a photon pair by using phonons in a manner as discussed by several authors~\cite{OMSEnt}. It should be kept in mind that although we produce a photon pair, the mediating mechanism is an active mechanism which puts a limit on the amount of achievable squeezing. This is in contrast to the situation with a downconverter where the crystal participates in a passive manner in the sense that it does not contribute to quantum noise. We show generation of very large squeezing even at temperatures like $10$mK. The large squeezing is a consequence of active phonon nonlinearities which become large due to the resonant nature of the underlying processes.

The organization of this paper is as follows: In Sec. II we present the basic model, underlying equations and the different parametric processes in the two cavities. In Sec. III we present the calculation of the squeezing spectra for the field which is a linear superposition of the two output fields. In Sec. IV we present numerical results for squeezing. We explain the origin of squeezing via the phonon mediated four-wave mixing (FWM) process. We also compare our squeezing scheme to those obtained using other methods. In Sec. V we investigate the effect of the Brownian noise of the mirror and its effect on the output state purity. In Sec. VI we present our conclusions.

\section{Model and fluctuating quantum fields}
As mentioned in the introduction, we consider a double cavity optomechanical system model, in which a mechanical resonator coated with perfect reflecting films on both sides are coupled to two identical optical cavities. One possible realization is using ``membrane-in-the-middle'' setup~\cite{Vitali}, except that we assume the membrane is fully reflective on both sides. We note that double cavity optomechanical systems have gained considerable importance because of many possible applications. These includes two-mode electromagnetically induced transparency~\cite{app1}, electromagnetically induced absorption~\cite{app2}, quantum state conversion~\cite{double}, optical wavelength conversion~\cite{app4}, enhancing quantum nonlinearities~\cite{app5}, controllable optical bistability~\cite{app6}, photon blockade~\cite{app7}, quantum-nondemolition measurement~\cite{app8} and entangled photon pair generation~\cite{OMSEnt}.
\begin{figure}[phtb]
 \includegraphics[width=0.26\textwidth]{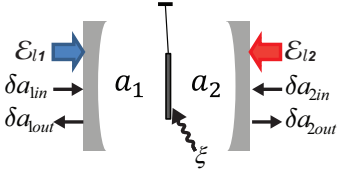}
 \caption{\label{fig1}{ Schematic of the proposed double cavity optomechanics where cavity 1 (2) fed by blue (red) detuned driving lasers and vacuum inputs are coupled to the same mechanical resonator mediated in thermal bath. $\mathcal{E}_{li}$, $a_i$, $\delta a_{i\text{in}}$ and $\delta a_{i\text{out}}$ denote the classical driving field, in-cavity optical field, input quantum vacuum noise and output quantum fluctuation for cavity $i$, respectively, and $\xi$ denotes the mechanical noise.  }}
\end{figure}

Let us denote the optical modes using the annihilation operators $a_i$ for the cavity $i$. We denote the mechanical mode of the resonator using the normalized displacement operator $Q=\sqrt{m\omega_m/\hbar}q$ and momentum operator $P=1/\sqrt{m\hbar \omega_m}p$ where $q$, $p$, $m$ and $\omega_m$ are the displacement, momentum,  mass and the oscillation frequency of the mirror, respectively. The interaction between the optical and mechanical modes arises from the radiation pressure of light, which results in a change in the cavity length and hence the cavity frequency. The radiation pressure interaction can be written as~\cite{GSAbook,review} $-(\hbar\omega_{ci}/L_i)q a_i^\dag a_i=-\hbar g_i Q a_i^\dag a_i$, where $\omega_{ci}$ and $L_i$ are the resonance frequency and length of the empty cavity $i$. We denote the coupling rate by $g_i=\frac{\omega_{ci}}{L_i}\sqrt{\frac{\hbar}{m\omega_m}}$. In order to enhance the optomechanical coupling, a driving laser with an amplitude $\mathcal{E}_{li}$ and frequency $\omega_{li}$ is applied to each cavity. We treat the driving lasers classically since they are strong. Then the Hamiltonian for the system can be written as
\begin{align} \label{1}
	H &= \sum_{i=1,2} [\hbar\omega_{ci} a_i^\dag a_i + i\hbar\mathcal{E}_{li}(a_i^\dag\mathrm{e}^{-i\omega_{li}t}-a_i\mathrm{e}^{i\omega_{li}t})] \nonumber \\
	& \quad + \frac12 \hbar\omega_m(Q^2+P^2)  - \hbar (g_1 a_1^\dag a_1 - g_2 a_2^\dag a_2) Q.
\end{align}
The driving laser amplitude is related to its power $\mathcal{P}_{li}$ by $\mathcal{E}_{li} = \sqrt{2\kappa_i\mathcal{P}_{li}/(\hbar\omega_{li})}$ and $2\kappa_i$ is the decay rate of the cavity $i$. It is convenient to rewrite the Hamiltonian into a new picture using transformation $\exp[-i\sum(\omega_{li}a_i^\dag a_i t)]$, then
\begin{align} \label{2}
	H &= \sum_{i=1,2} [\hbar(\omega_{ci}-\omega_{li})b_i^\dag b_i + i\hbar\mathcal{E}_{li}(b_i^\dag-b_i)] \nonumber \\
	& \quad + \frac12 \hbar\omega_m(Q^2+P^2)  - \hbar (g_1 b_1^\dag b_1 - g_2 b_2^\dag b_2) Q,
\end{align}
with $b_i$'s defined by $a_i=b_i\mathrm{e}^{-i\omega_{li}t}$. The quantum Langevin equations governing the system can be obtained by using the Heisenberg equations and adding the corresponding decay and input noise terms,
\begin{equation} \label{3}
	\begin{aligned}
		\dot{Q} &= \omega_m P, \\
		\dot{P} &= g(b_1^\dag b_1-b_2^\dag b_2) - \omega_m Q -\gamma_m P + \xi, \\
		\dot{b_1} &= -i(\omega_{c1}-\omega_{l1} - gQ)b_1 - \kappa_1 b_1 + \mathcal{E}_{l1} + \sqrt{2\kappa_1}b_{1\text{in}}, \\
		\dot{b_2} &= -i(\omega_{c2}-\omega_{l2} + gQ)b_2 - \kappa_2 b_2 + \mathcal{E}_{l2} + \sqrt{2\kappa_2}b_{2\text{in}}, \\		
	\end{aligned}
\end{equation}
where $b_{1\text{in}}$ and $b_{2\text{in}}$ are the input vacuum noise with correlation fluctuations
\begin{equation}\label{3a}
  \langle b_{i\text{in}}(t)b_{j\text{in}}^\dag(t')\rangle = \delta_{ij} \delta(t-t').
\end{equation}
The noise term $\xi$ stems from the thermal noise of the mechanical resonator at a finite temperature $T$, which obeys
\begin{equation*}
  \langle \xi(t)\xi(t') \rangle = \frac{\gamma_m}{2\pi\omega_m}\int\omega\mathrm{e}^{-i\omega(t-t')} \left[1+\coth\left(\frac{\hbar\omega}{2K_BT}\right)\right] \mathrm{d}\omega,
\end{equation*}
where $K_B$ is the Boltzmann constant. We follow the standard procedure and solve Eq.~(\ref{3}) perturbatively by separating the classical mean value and the fluctuation of each operator,
\begin{equation} \label{4}
	Q=Q_s+\delta Q, \qquad P=P_s+\delta P, \qquad b_i=b_{is} + \delta b_i,
\end{equation}
for $i=1,2$. In this way, we can solve for the classical mean values of the optical fields as $b_{is}=\frac{\mathcal{E}_{li}}{\kappa_i+i\Delta_i}$ and $Q_s=(|b_{1s}|^2g_1-|b_{2s}|^2g_2)/\omega_m$ where $\Delta_i=\omega_{ci}-\omega_{li}\mp g_iQ_s$ are the mean detuning of the cavities and $-$ is for $i=1$ and $+$ for $i=2$. Then the linearized quantum Langevin equations are given by
\begin{equation}\label{5}
    \begin{aligned}
        \delta\dot{Q} & =\omega_{m}\delta P, \\
		\delta\dot{P} & =g_1(b_{1s}^*\delta b_1+b_{1s}\delta b_1^\dag) - g_2(b_{2s}^*\delta b_2+b_{2s}\delta b_2^\dag) \\
            & \qquad -\omega_{m}\delta Q-\gamma_{m}\delta P + \xi, \\
		\delta\dot{b}_{1} & = -(\kappa_1 + i\Delta_1)\delta b_1 + ig_1 b_{1s}\delta Q + \sqrt{2\kappa_1} b_{1\text{in}}, 	\\
		\delta\dot{b}_{2} & = -(\kappa_2 + i\Delta_2)\delta b_1 - ig_2 b_{2s}\delta Q + \sqrt{2\kappa_2} b_{2\text{in}},
    \end{aligned}
\end{equation}
It is convenient to work with the new optical and mechanical annihilation operators defined as
\begin{equation}\label{6}
\begin{aligned}
  \tilde{b}_1 &= \delta b_1\mathrm{e}^{-i\omega_m t}, \quad  & c\mathrm{e}^{-i\omega_m t} &=  (\delta Q + \delta P)/\sqrt{2}, \\
  \tilde{b}_2 &= \delta b_2\mathrm{e}^{i\omega_m t},   & c^\dag\mathrm{e}^{i\omega_m t} &=  (\delta Q - \delta P)/(\sqrt{2}i),
\end{aligned}
\end{equation}
and the input field fluctuations defnied as $\tilde{b}_{i\text{in}} = \delta b_{i\text{in}}\mathrm{e}^{\mp i\omega_m t}$, with $\mp$ for $i=1,2$.
These operators obey the equations
\begin{equation}\label{inc7}
\begin{aligned}
  \dot{c} &=  - \frac{\gamma_m}{2}(c - c^\dag\mathrm{e}^{2i\omega_m t}) + f(t) \\
   &\qquad + i\frac{g_1}{\sqrt{2}}(b_{1s}^*\tilde{b}_1\mathrm{e}^{2i\omega_m t} + b_{1s}\tilde{b}_1^\dag)  \\
   &\qquad - i \frac{g_2}{\sqrt{2}}(b_{2s}^*\tilde{b}_2 + b_{2s}\tilde{b}_2^\dag\mathrm{e}^{2i\omega_m t}),  \\
  \dot{\tilde{b}}_1  &=  -(\kappa_1 + i\Delta_1 + i\omega_m) b_1 + \sqrt{2\kappa_1} \tilde{b}_\text{1\text{in}}  \\
   &\qquad + i\frac{g_1}{\sqrt{2}} b_{1s}(c\mathrm{e}^{-2i\omega_m t} + c^\dag), 	\\
  \dot{\tilde{b}}_2  &=  -(\kappa_2 + i\Delta_2 - i\omega_m) b_2 + \sqrt{2\kappa_2} \tilde{b}_\text{2\text{in}} \\
   &\qquad - i\frac{g_2}{\sqrt{2}} b_{2s}(c + c^\dag\mathrm{e}^{2i\omega_m t}).
\end{aligned}
\end{equation}
The rapidly rotating terms in (\ref{inc7}) correspond to nonresonant FWM processes, for example in cavity $2$ a red pump $\omega_c-\omega_m$ and a probe $\omega_c$ can produce a photon of frequency $\omega_c-2\omega_m$ and another photon of frequency $\omega_c$. The generation at $\omega_c$ is a resonant process whereas the generation at $\omega_c-2\omega_m$ is a nonresonant process. We drop all the nonresonant processes. Thus by dropping the rapidly rotating terms at frequencies $2\omega_m$, we obtain
\begin{equation}\label{7}
\begin{aligned}
  \dot{c} &=  - \frac{\gamma_m}{2}c + iG_1\tilde{b}_1^\dag - i G_2^*\tilde{b}_2 + f(t),  \\
  \dot{\tilde{b}}_1^\dag &=  -(\kappa_1 - ix_1)\tilde{b}_1^\dag - iG_1^*c + \sqrt{2\kappa_1} \tilde{b}^\dag_\text{1\text{in}}, \\
  \dot{\tilde{b}}_2 &=  -(\kappa_2 + ix_2)\tilde{b}_2 - iG_2c + \sqrt{2\kappa_2} \tilde{b}_\text{2\text{in}},
\end{aligned}
\end{equation}
where $x_1=\Delta_1 + \omega_m$, $x_2=\Delta_2 - \omega_m$ and $G_i=b_{is}g_i/\sqrt{2}$ for $i=1,2$. Notice that $G_i$ is a pure imaginary number by the definition of $b_{is}$ under the resolved side-band regime, $\Delta\gg\kappa_i$. Since the coupling laser in cavity $1(2)$ is blue(red) detuned by an amount $\omega_m$, $x_1\sim x_2\sim0$. $f(t)$ has the correlation relation
\begin{equation}\label{8}
\begin{aligned}
  \langle f(t)f^\dag(t')\rangle &= \gamma_{m} (\bar{n}_\text{th}+1)\delta(t-t'), \\
  \langle f^\dag(t)f(t')\rangle &= \gamma_{m} \bar{n}_\text{th}\delta(t-t'),
\end{aligned}
\end{equation}
where $\bar{n}_\text{th}(\omega)= [\exp(\hbar\omega/K_B T)-1)]^{-1}$ is the mean phonon number at temperature $T$. In order to solve these equations, we transform them into the frequency domain using $A(t)=\frac{1}{2\pi} \int_{-\infty}^{+\infty} A(\omega)\mathrm{e}^{-i\omega t} \mathrm{d}\omega$, and $A^\dag(t)=\frac{1}{2\pi} \int_{-\infty}^{+\infty} A^\dag(-\omega)\mathrm{e}^{-i\omega t} \mathrm{d}\omega$, so that $A^\dag(-\omega)=[A(-\omega)]^\dag$. Then the correlation relation is
\begin{equation}\label{8in}
\begin{aligned}
  & \langle \tilde{b}_i(\omega)\tilde{b}_j^\dag(-\omega')\rangle=2\pi\delta_{ij}\delta(\omega+\omega'), \\
  & \langle f^\dag(t)f(t')\rangle = 2\pi\gamma_{m} \bar{n}_\text{th}\delta(\omega+\omega'), \\
  & \langle f(t)f^\dag(t')\rangle = 2\pi\gamma_{m} (\bar{n}_\text{th}+1)\delta(\omega+\omega').
\end{aligned}
\end{equation}
Using the input-output relations $\tilde{b}_{i\text{out}}=\sqrt{2\kappa_i}\tilde{b}_i - \tilde{b}_{i\text{in}}$, the output optical fields can be calculated as
\begin{align}
  \tilde{b}_{1\text{out}}(\omega) &= E_1(\omega) \tilde{b}_{1\text{in}}(\omega) + F_1(\omega) \tilde{b}^\dag_{2\text{in}}(-\omega) \nonumber \\
   & \qquad + V_1(\omega) f^\dag(-\omega), \label{9} \\
  \tilde{b}_{2\text{out}}(\omega) &= E_2(\omega) \tilde{b}_{2\text{in}}(\omega) + F_2(\omega) \tilde{b}^\dag_{1\text{in}}(-\omega) \nonumber \\
   & \qquad + V_2(\omega) f(\omega), \label{10}
\end{align}
where
\begin{equation}\label{11}
\begin{aligned}
  E_1(\omega) &= \frac{2\kappa_1}{D^*(-\omega)}\frac{|G_1|^2}{(\kappa_1+ix_1+i\omega)^2} + \frac{2\kappa_1}{\kappa_1+ix_1+i\omega} -1, \\
  F_1(\omega) &= -\frac{\sqrt{2\kappa_12\kappa_2}G_1^*G_2^*}{D^*(-\omega)(\kappa_1+ix_1+i\omega)(\kappa_2-ix_2-i\omega)}, \\
  E_2(\omega) &= -\frac{2\kappa_2}{D(\omega)}\frac{|G_2|^2}{(\kappa_2+ix_2-i\omega)^2} + \frac{2\kappa_2}{\kappa_2+ix_2-i\omega} -1, \\
  F_2(\omega) &= \frac{\sqrt{2\kappa_12\kappa_2}G_1G_2}{D(\omega)(\kappa_1-ix_1+i\omega)(\kappa_2+ix_2-i\omega)}, \\
  V_1(\omega) &= -\frac{iG_1\sqrt{2\kappa_1}}{D^*(-\omega)(\kappa_1+ix_1+i\omega)}, \\
  V_2(\omega) &= -\frac{iG_2\sqrt{2\kappa_2}}{D(\omega)(\kappa_2+ix_2-i\omega)}, \\
  D(\omega) &= -\frac{|G_1|^2}{\kappa_1-ix_1+i\omega} + \frac{|G_2|^2}{\kappa_2+ix_2-i\omega} + (\frac{\gamma_m}{2}-i\omega).
\end{aligned}
\end{equation}
The observed fields $\delta a_{i\text{out}}(t)$ are related to $\tilde{b}_{i\text{out}}(t)$ via
\begin{align}
  \delta a_{1\text{out}} &= \delta b_{1\text{out}} \mathrm{e}^{-i\omega_{l1} t} = \tilde{b}_{1\text{out}} \mathrm{e}^{-i(\omega_{l1}-\omega_m) t} = \tilde{b}_{1\text{out}} \mathrm{e}^{-i \omega_c t}, \label{11a}\\
  \delta a_{2\text{out}} &= \delta b_{2\text{out}} \mathrm{e}^{-i\omega_{l1} t} = \tilde{b}_{2\text{out}} \mathrm{e}^{-i(\omega_{l2}+\omega_m) t} = \tilde{b}_{2\text{out}} \mathrm{e}^{-i \omega_c t}. \label{11b}
\end{align}
Hence, we see that they have the simple relation $\delta a_{i\text{out}}=\tilde{b}_{i\text{out}} \mathrm{e}^{-i \omega_c t}$. The above equations are valid under the condition that the system is in the stable regime. The sufficient and necessary condition for stability is that the coefficient matrix of the differential equations (\ref{7}) by dropping fluctuating forces must have eigenvalues $\lambda$ with negative real part,
\begin{equation}\label{14a}
    \begin{vmatrix}
        -\gamma_m/2-\lambda & iG_1 & -iG_2^* \\
        -iG_1^* & -(\kappa_1-ix_1)-\lambda & 0 \\
        -iG_2 & 0 & -(\kappa_2+ix_2)-\lambda
    \end{vmatrix}
    =0.
\end{equation}
By using the Routh-Hurwitz criterion~\cite{Hurwitz},  we get the stability condition $|G_1|^2/\kappa_1 - |G_2|^2/\kappa_2<\gamma_m/2$ when $x_1\sim x_2\sim0$. If this condition is violated, the system goes into the instable regime.

\begin{figure}[phtb]
 \includegraphics[width=0.48\textwidth]{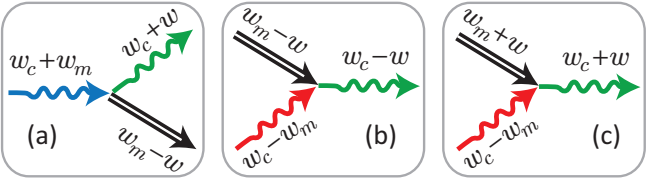}
 \caption{\label{fig8}{The photon-phonon interaction processes in cavity 1 \textbf{(a)} and \textbf{(c)} and in cavity 2 \textbf{(b)}. }}
\end{figure}

We now give the meaning of the coefficients $E_i$, $F_i$ and $V_i$ in (\ref{9}) and (\ref{10}). These coefficients are obtained to all orders in the strengths of the blue and red pumps. The $E_i$'s and $F_i$'s to second order in $G_i$ can be given simple physical interpretations. Let us first consider an incoming vacuum photon from cavity $1$. It should be borne in mind that the frequency $\omega$ from the cavities corresponds to $\omega_c+\omega$ according to Eqs.~(\ref{11a}) and (\ref{11b}). This produces a vacuum photon of frequency $\omega_c+\omega$ in cavity $1$ and a photon of frequency $\omega_c-\omega$ in cavity $2$. The reason for the production of a photon of frequency $\omega_c-\omega$ can be understood as follows: A blue detuned photon of frequency $\omega_c+\omega_m$ produces a phonon of frequency $\omega_m-\omega$ and a photon of frequency $\omega_c+\omega$. The phonon of frequency $\omega_m-\omega$ interacts with the red detuned pump of frequency $\omega_c-\omega_m$. This is shown in the diagram in Fig~\ref{fig8}.
The term $F_2(-\omega)$ in Eq.~(\ref{10}) represents the combined effect of these two processes. We can similarly understand $F_1(-\omega)$ in Eq.~(\ref{9}) by considering an incoming vacuum photon from cavity $2$. Note that these are the diagrams which contribute to the lowest order in $G_1G_2$ in the expression for $F_2(-\omega)$. The term proportional to $|G_1|^2$ in $E_1(\omega)$ arises from the diagram Fig.~\ref{fig8}(a). The $V_i$ terms in (\ref{9}) and (\ref{10}) correspond to the quantum noise which is added by either the thermal phonons or vacuum phonons. Note that in lowest order in $G_i$'s, we can interpret the last term in (\ref{10}) by saying that a thermal phonon of frequency $\omega_m+\omega$ combined with a red photon of frequency $\omega_c-\omega_m$ to produce a photon of frequency $\omega_c+\omega$ as shown in the Fig.~\ref{fig8}(c). Similarly in Eq.~(\ref{10}) a thermal phonon or a vacuum phonon of frequency $\omega_m-\omega$ and a photon of frequency $\omega_c+\omega$ combine to create a blue photon $\omega_c+\omega_m$. This is the reverse of the process shown in Fig.~\ref{fig8}(a).

\section{Squeezing spectra}
\begin{figure}[bpht]
 \includegraphics[width=0.2\textwidth]{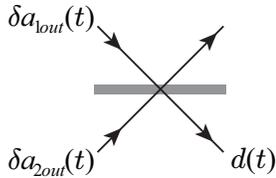}
 \caption{\label{fig2}{ The combination of the output fields $\delta a_{i\text{out}}$'s from two cavities using a $50/50$ beam splitter. }}
\end{figure}

For studying the squeezing spectra, we combine the output fields $\delta a_{1\text{out}}$ and $\delta a_{2\text{out}}$ to construct the field $d$ as shown in Fig.~\ref{fig2}.
To make it more general, we add a phase difference $\theta$ between the output fields, then $d(t)$ can be written as
\begin{align}\label{12a}
  d(t) &= \frac{1}{\sqrt{2}} [\delta a_{1\text{out}}(t) + \mathrm{e}^{i\theta}\delta a_{2\text{out}}(t)] \nonumber \\
  &= \frac{1}{\sqrt{2}} [\tilde{b}_{1\text{out}}(t) + \mathrm{e}^{i\theta}\tilde{b}_{2\text{out}}(t)]\mathrm{e}^{-i \omega_c t}.
\end{align}
In the frame rotating with the cavity frequency $\omega_c$,
\begin{equation}\label{12b}
  \tilde{d}(t) = d(t) \mathrm{e}^{i \omega_c t} = \frac{1}{\sqrt{2}} [\tilde{b}_{1\text{out}}(t) + \mathrm{e}^{i\theta}\tilde{b}_{2\text{out}}(t)],
\end{equation}
which obeys the commutation relation $[\tilde{d}(t), \tilde{d}^\dag(t')] = \delta(t-t')$. The spectrum of $\tilde{d}(t)$ can be calculated using Eqs.~(\ref{9})-(\ref{11}). We define as usual the quadrature variable $X_\phi(t) = [\tilde{d}(t)\mathrm{e}^{-i\phi} + \tilde{d}^\dag(t)\mathrm{e}^{i\phi}]/\sqrt{2}$, and hence in the frequency domain
\begin{align}\label{12}
  X_\phi(\omega) &= \frac{1}{\sqrt{2}} [\tilde{d}(\omega)\mathrm{e}^{-i\phi} + \tilde{d}^\dag(-\omega)\mathrm{e}^{i\phi}] \nonumber \\
  &= \frac{1}{2} \Big[\Big(\tilde{b}_{1\text{out}}(\omega) + \mathrm{e}^{i\theta}\tilde{b}_{2\text{out}}(\omega)\Big)\mathrm{e}^{-i\phi} \nonumber \\
  & \qquad\quad + \Big(\tilde{b}^\dag_{1\text{out}}(-\omega) + \mathrm{e}^{i\theta}\tilde{b}^\dag_{2\text{out}}(-\omega)\Big)\mathrm{e}^{i\phi}\Big] \nonumber \\
  &= \frac12 \Big[E(\omega)\tilde{b}_{1\text{in}}(\omega) + E^*(-\omega)\tilde{b}^\dag_{1\text{in}}(-\omega)   \nonumber \\
  & \qquad\quad + F(\omega) \tilde{b}_{2\text{in}}(\omega) + F^*(-\omega) \tilde{b}^\dag_{2\text{in}}(-\omega) \nonumber \\
  & \qquad\quad + V(\omega) f(\omega)+ V^*(-\omega) f^\dag(-\omega) \Big],
\end{align}
where
\begin{align}\label{13}
  E(\omega) &= E_1(\omega)\mathrm{e}^{-i\phi} + F_2^*(-\omega)\mathrm{e}^{i\phi-i\theta},  \nonumber \\
  F(\omega) &= E_2(\omega)\mathrm{e}^{i\theta-i\phi} + F_1^*(-\omega)\mathrm{e}^{i\phi},  \nonumber \\
  V(\omega) &= V_1(\omega)\mathrm{e}^{-i\phi} + V_2^*(-\omega)\mathrm{e}^{i\phi-i\theta}.
\end{align}
The squeezing spectrum defined as $\langle X_\phi(\omega)X_\phi(\omega') \rangle = 2\pi S_\phi(\omega)\delta(\omega+\omega')$ can then be obtained using the correlation relations (\ref{3a}) and (\ref{8})
\begin{align}\label{14}
  S_\phi(\omega) =& \frac{1}{2\pi} \int \langle X_\phi(\omega)X_\phi(\omega') \rangle \mathrm{d}\omega' \nonumber \\
  =& \frac14  \Big[ |V(\omega)|^2 \gamma_m (\bar{n}_\text{th}+1) + |V(-\omega)|^2 \gamma_m \bar{n}_\text{th}  \nonumber \\
  &\quad + |E(\omega)|^2 + |F(\omega)|^2 \Big] .
\end{align}
We note that if $\tilde{d}(t)$ were a vacuum field, then
\begin{equation}\label{14inc}
  S_\phi(\omega)=1/2
\end{equation}
Hence we define the normalized squeezed parameter as $2S_\phi(\omega)$. The magnitude of squeezing in dB units is then $-10\log_{10}(2S_\phi)$.

\section{Squeezing in the output fields from a double cavity optomechanics}
We have studied the physics of the squeezing process in optomechanics in analogy to the down conversion process, and we expect Eq.~(\ref{14}) to yield squeezing. It is convenient to introduce the cooperativity parameter $C_i=2|G_i|^2/(\kappa_i\gamma_m)$ for each cavity $i$.
\begin{figure}[phtb]
 \includegraphics[width=0.49\textwidth]{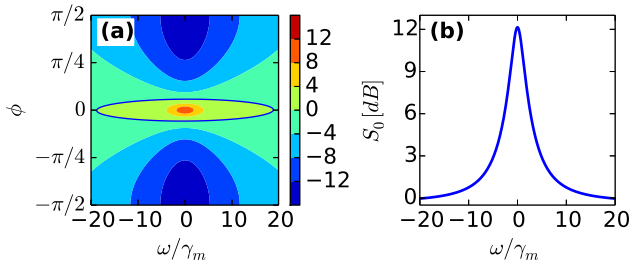}
 \caption{\label{fig4}{(a) The density plot of the quadratures of the output field $b(\omega)$ with $-\Delta_1= \Delta_2= \omega_m$ at zero temperature. The middle region between the thick contours is squeezed. (b) The field amplitude ($\phi=0$) quadrature. The parameter set used in the plots are $\omega_m=2\pi\times 50$MHz, $\kappa=2\pi\times1$MHz, $\gamma_m=2\pi\times1$kHz, $G_2=i 2\pi\times0.1$MHz ($C_2=20$), $G_1=-G_2/\sqrt{2}$ ($C_1=10$), $x_1=x_2=0$ and $\theta=\pi$. }}
\end{figure}
We illustrate the features of our expected variance in the quadratures of the output field $d(t)$ in Fig.~\ref{fig4}(a) with $\theta=\pi$ and for cooperativities $C_2=2C_1=20$. In the plot, we set $\kappa_1=\kappa_2=\kappa$. The complete set of parameters are given in the caption. To create this map, we used Eq.~(\ref{14}) at zero temperature. In the diagram, we observe the largest magnitude of squeezing in the amplitude quadrature $S_0$ (see Fig.~\ref{fig4}(b)).  The magnitude of squeezing at $\omega=0$ is about $12$dB. As one rotates towards the phase quadrature $S_{\pi/2}$, the squeezing magnitude decreases and it eventually turns into antisqueezing.

It turns out that one can find the value of $S_\varphi$ at $\omega=0$ analytically if we use the approximation $\gamma,\kappa_i\ll\omega_m$, then
\begin{equation}\label{15}
    S_0(0) = \frac{1+(\sqrt{C_1}-\sqrt{C_2})^4}{2(1-C_1+C_2)^2} + \frac{(\sqrt{C_1}-\sqrt{C_2})^2(2\bar{n}_\text{th}+1)}{(1-C_1+C_2)^2}.
\end{equation}
The first term in Eq.~(\ref{15}) describes the noise squeezing of the input vacuum field and the second term arises from the noise due to the thermal bath phonons. We concentrate on the first term for now and discuss the optomechanical squeezing effects. A short derivation shows that $S_0(0)$ approaches to its minimum value $S_\text{min}= \frac{1}{4C_2+2}$ when $\frac{C_1}{C_2} \to [\frac{(2C_2+1)-\sqrt{4C_2+1}}{2C_2} ]^2$ and ignoring the mechanical noise. For $C_i\gg1$, they can be approximated as $S_\text{min}= 1/(4C_2)$ when $\frac{C_1}{C_2} \to [1-\frac{1}{\sqrt{C_2}}]^2$. Thus the squeezing reaches its maximum magnitude. Nonetheless, one also has to keep in mind of the stability condition from Eq.~(\ref{14a}). Routh-Hurwitz stability criterion imposes the condition for the enhanced coupling rates into a dynamically stable range $C_1-C_2<1$. In the resolved side-band limit and large cooperativity limit, it requires $C_1<C_2$. Therefore, we can draw the conclusion that the effect of squeezing is optimized as $C_1/C_2$ approaches from a small value to $[1-\frac{1}{\sqrt{C_2}}]^2$, which is close to the limit when instability occurs. This result is analogous, for example, to the result in optical bistability that squeezing increases as one approaches the instability point.

\begin{figure}[phtb]
 \includegraphics[width=0.49\textwidth]{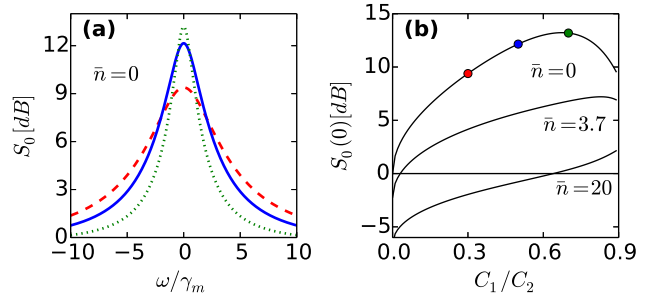}
 \caption{\label{fig5}{(a) The field amplitude ($\phi=0$) quadrature with $C_1/C_2=0.7$(red dashed), $=0.5$(blue solid) and $=0.3$(green dotted) at $T=0$. (b) The squeezing magnitude at $\omega=0$ versus $C_1/C_2$ by fixing $C_2=20$ at different temperatures. The three dots on the top curve corresponds to different curve in (a)  Other parameters  are the same as in Fig.~\ref{fig4}.  }}
\end{figure}

We illustrate the dependence of the squeezing magnitude on the cooperativity parameter or on the pump power in Fig.~\ref{fig5}.
Fig.~\ref{fig5}(a) shows the squeezing spectrum under different ratios of $C_1/C_2$ and we see that squeezing spectrum gains magnitude but loses width when $C_1/C_2$ increases from $0.3$(dotted) to $0.5$(solid) and $0.7$(dashed). In Fig.~\ref{fig5}(b), we plot the squeezing magnitude at $\omega=0$ as a function of $C_1/C_2$ when the temperatures are both zero and nonzero. We see that when $C_1=0$, the vacuum optical inputs only interact with cavity $2$, and no squeezing process is undergoing. The incoherent phonons from the mirror in the thermal bath result in fluctuations in the optical output field, hence $S_0(0) \leq0$ when $T\geq0$. The magnitude of squeezing $S_0(0)$ increases with increasing $C_1/C_2$ until it reaches the maximum squeezing. At $T=0$, the maximum squeezing occurs at $C_1/C_2 = 0.7 \approx 1-2/\sqrt{C_2}$ and $C_2=20$. The system loses squeezing magnitude after this point if $C_1/C_2$ keeps increasing. It should be borne in mind that for $C_1/C_2 \to 1$, the system approaches to the threshold for instability. When it is too close to the threshold, the linearization procedure used to calculate fluctuations begins to break down.

\begin{figure}[phtb]
 \includegraphics[width=0.28\textwidth]{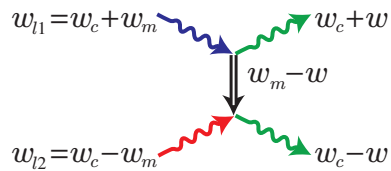}
 \caption{\label{fig14}{Generation of squeezed states via phonons in FWM process.}}
\end{figure}

The physics in the generation of the squeezed vacuum states can be interpreted using the FWM process via phonons, as shown in Fig.~\ref{fig14}.
In cavity $1$, A blue detuned driving laser photon ($\omega_{l1}=\omega_c+\omega_m$), when being scattered by the mechanical oscillator, produces a phonon ($\omega_m-\omega$) and a photon at a lower frequency $\omega_c+\omega$. At the same time in cavity $2$, a red detuned driving laser photon ($\omega_{l2}=\omega_c-\omega_m$) by absorbing the phonon ($\omega_m-\omega$) produces a photon at $\omega_c-\omega$. These processes are resonantly enhanced if both the generated photons are close to the cavity resonance frequency. Equivalently, the physics can be described by the effective Hamiltonian for the FWM process in Fig.~\ref{fig14}: $\displaystyle a_{l1}a_{l2}\int\Phi(\omega) a^\dag_{\omega_c-\omega} a^\dag_{\omega_c+\omega}\mathrm{d}\omega + h.c. \approx \alpha \int\Phi(\omega) a^\dag_{\omega_c-\omega} a^\dag_{\omega_c+\omega}\mathrm{d}\omega + h.c.$, when the strong driving lasers $a_{l1},a_{l2}$ can be approximated classically by a number $\alpha$. Here $\Phi(\omega)$ depends on the details of the optomechanical cavities. Such an interaction has been extensively studied in quantum optics~\cite{Wolf,Wallsbook,GSAbook} and is known to lead to the generation of quantum entanglement as well as quantum squeezing. In the context of the double cavity optomechanics, the generation of entangled pairs has been discussed previously~\cite{OMSEnt}.

Our double cavity optomechanics proposal is fundamentally different from the aforementioned ponderomotive squeezing~\cite{Heidmann,squeezing,Brooks,PRX} which has been experimentally realized by Purdy et al~\cite{PRX} in a membrane setup and by Safavi-Naeini et al~\cite{squeezing} in a waveguide-coupled zipper optomechanical cavity. In their experiments, a coherent input at the cavity resonance frequency is applied and the quantum noise of coherent light is reduced by using radiation pressure to push the mechanical vibrating membrane that, in turn, feeds back on the light's phase. The output squeezed light is generated at the side-band of the cavity frequency detuned by $\omega_m$, which is approximately equal to the cavity linewidth. The degree of noise reduction depends on the optomechanical coupling strength. They did not use side-band resolved condition and reported reasonable squeezing (several dB) under experimental conditions. We work in the side-band resolved limit and by using two different parametric processes, where the driving lasers are red and blue detuned, produce photon pairs. Such photon pairs are then combined with a beam splitter to produce squeezing. As a beneficial of this particular driving manner, the squeezed output fields are on resonance to the cavity frequency and hence can be made strong. The red detuned driving field, on the other hand, inherently ensures the stability without requiring any extra cooling laser as long as the red detuned pump interaction is stronger than the blue detuned one.

\section{Effect of the Brownian noise of the mirror on squeezing}
It is known that the squeezing is degraded by any kinds of noise effects.
\begin{figure}[hptb]
 \includegraphics[width=0.49\textwidth]{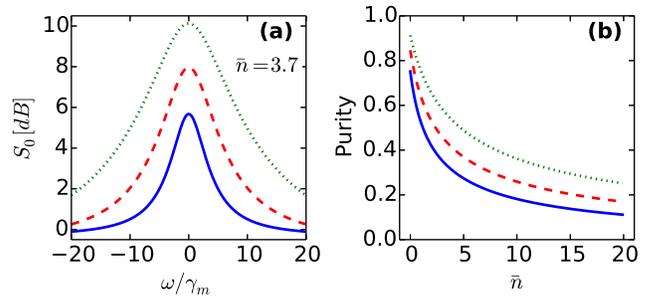}
 \caption{\label{fig7}{(a) The spectrum of amplitude quadrature ($\phi=0$) at $T=10$mK correspondingly $\bar{n}_\text{th}=3.7$; and (b) purity of the squeezing state with fixed $C_2/C_1=1/2$ and changing $C_2=20$(blue solid), $=40$(red dashed) and $=80$(green dotted). Other parameters are the same as in Fig.~\ref{fig4}.  }}
\end{figure}
In optomechanics, the Brownian noise of the mirror makes the observation of quantum effects difficult. As we analyzed in the last section, the squeezing mechanism in our scheme relies on the coherent phonons generated by the driving field to actively transfer quantum coherence between two cavity fields. However, at the same time, the mirror is mediated in the thermal reservoir which excites incoherent phonons and hence limits the purity of the squeezed fields. At a high temperature, the system even loses the squeezing ability. This is illustrated in Fig.~\ref{fig5}(b), where the curve for $\bar{n}=20$ shows antisqueezing when $C_1/C_2<0.6$.

We now investigate the effect of the $V$ terms in Eq.~(\ref{12}) on the possible amounts of squeezing. In Fig.~\ref{fig7}(a), we plot the output field amplitude quadrature at a finite temperature $T=10$mK and correspondingly $\bar{n}_\text{th}=3.5$. Comparing the solid curves in Fig.~\ref{fig5}(a) and in Fig.~\ref{fig7}(a) which are both plotted using the same parameters other than different bath temperatures, one can clearly see the squeezing magnitude decreases from $12$ to $6$ when the temperature increases from $0$ to $10$mK. Interestingly, in our system, the decreasing of squeezing due to rising the bath temperature can be compensated by increasing the cooperativity, in way of enhancing the coupling constant or reducing the decaying rates. Now we concentrate on Fig.~\ref{fig7}(a). When $C_2$ is increased from $20$ (solid) to $40$ (dashed) and to $80$ (dotdashed), the squeezing magnitude increases from $6$dB to $8$dB and $10$dB, successively. The widths of the squeezed spectrum are increased as well. This agrees with Eq.~(\ref{15}) from which we find increasing $C_2$ essentially reduces the effect of $\bar{n}$. The equation (\ref{15}) also suggests that a larger cooperativity is always preferable in order to generate a large squeezing magnitude at nonzero temperatures, although it never exceeds the zero temperature case. One has squeezing as long as the right hand side of Eq.(\ref{15}) is less than $0.5$. For $C_2=2C_1=20$, $S_0(\omega)$ has the value $0.030+0.028\bar{n}_\text{th}$.

We are discussing the spectrum of squeezing. It is also interesting to examine the purity of the output state which is given in terms of the density matrix $\rho$ by the condition $\mathrm{Tr}\rho^2 =1$. We are dealing with a Gaussian Wigner function as Eqs.~(\ref{9})-(\ref{10}) are linear in vacuum operators. For a single mode of the field, the purity~\cite{purity} is defined by $\mathrm{Tr}\rho^2 = 1/\sqrt{\mathrm{det}\bold{\sigma}}$ where $\bold{\sigma}$ is the covariance matrix of the state
\begin{equation}\label{16}
  \bold{\sigma} = \begin{pmatrix}
    2\langle X_0^2\rangle & \langle X_0X_{\pi/2}+X_0X_{\pi/2}\rangle \\
    \langle X_0X_{\pi/2}+X_0X_{\pi/2}\rangle & 2\langle X_{\pi/2}^2\rangle
  \end{pmatrix},
\end{equation}
when $\langle X_0\rangle = \langle X_{\pi/2}\rangle$ which are zero under vacuum inputs. For our system, different frequency modes of the output field are uncorrelated as can be seen from Eqs.~(\ref{9})-(\ref{10}) and the properties in Eq.~(\ref{8in}) of the incoming vacuum fields and the mechanical Brownian noise. Hence, we can effectively use the result for the single mode. In Fig.~\ref{fig7}(b), we plot the purity of the output squeezed state at $\omega=0$ for different $C_2$'s. We see that the purity is close to unity at zero temperature. Note that at $T=0$, the vacuum fluctuations of the mechanical oscillator contribute to the non-purity of the output state. This can be seen from Eq.~(\ref{9}) which involves both $f$ and $f^\dag$ terms. As the temperature increases, the state purity decreases monotonically . The curves also show that the system with a higher cooperativity $C_2=80$ (top dotdshaed curve) loses purity slower than one with a lower cooperativity  $C_2=20$ (bottom solid curve).

\section{Conclusions}
In conclusion, we have shown how squeezing of the order of $10$dB can be generated in a double cavity optomechanical system. We searched for conditions under which the double cavity optomechanical system would lead to the generation of the photon pairs. We show that such a photon pair generation can be described by an effective interaction which are used for generating squeezing using parametric down conversion and four-wave mixing. However, there is one significant different: we generate photon pairs by using active participation of phonons. The phonon mediated processes lead to additional noise terms which degrade squeezing. The purity of the generated squeezed light is stronger with a large cooperativity in the resolved side-band regime. In light of the recent progress in optomechanics experiments that $\kappa\ll\omega_m$ and large cooperativity are realized in Ref.~\cite{Vitali}, our proposal shows very promising experimental feasibility.


\begin{thebibliography}{1}

\bibitem{Girvin11}  A. Nunnenkamp, K. B{\o}rkje, and S. M. Girvin, Phys. Rev. Lett. {\bf 107}, 063602 (2011).
\bibitem{Marquardt}  A. Kronwald and F. Marquardt, Phys. Rev. Lett. {\bf 111}, 133601 (2013).
\bibitem{Heidmann}  A. Heidmann, Y. Hadjar, M. Pinard, Appl. Phys. B {\bf 64}, 173(1997).
\bibitem{double} L. Tian, Phys. Rev. Lett. {\bf 108}, 153604 (2012); Y. -D. Wang and A. A. Clerk, Phys. Rev. Lett. {\bf 108}, 153603 (2012); T. A. Palomaki, J. W. Harlow, J. D. Teufel, R. W. Simmonds, and K. W. Lehnert, Nature {\bf 495}, 210 (2013); R. W. Andrews et al, arXiv:1310.5276v1 (2013).
\bibitem{OMSEnt} C. Genes, A. Mari, D. Vitali, and P. Tombesi, Adv. At., Mol. Opt. Phys. {\bf 57}, 33 (2009); Y.-D. Wang and A. A. Clerk, Phys. Rev. Lett. {\bf 110}, 253601 (2013); L. Tian, Phys. Rev. Lett. {\bf 110}, 233602 (2013); Z.-Q. Yin, and Y.-J. Han, Phys. Rev. A {\bf 79}, 024301 (2009); M. C. Kuzyk, S. J. van Enk, and H. Wang, Phys. Rev. A {\bf 88}, 062341 (2013).
\bibitem{Li}  J. Li, S. Gr\"{o}blacher, and M. Paternostro, New J. Phys. {\bf 15}, 033023 (2013).
\bibitem{sqmirror}  S. Huang and G. S. Agarwal, Phys. Rev. A {\bf 82}, 033811 (2010);  T. Weiss, C. Bruder, A. Nunnenkamp, New J. Phys. {\bf 15}, 045017 (2013).
\bibitem{Stamper-Kurn}  T. Botter, D. W. C. Brooks, N. Brahms, S. Schreppler, Dan M. Stamper-Kurn, Phys. Rev. A {\bf 85}, 013812 (2012).
\bibitem{Gong}  Y. -C. Liu, Y. -F. Xiao, Y. -L. Chen, X. -C. Yu, and Q. Gong, Phys. Rev. Lett. {\bf 111}, 083601 (2013).
\bibitem{Clerk}  M. -A. Lemonde, N. Didier, and A. A. Clerk, Phys. Rev. Lett. {\bf 111}, 053602 (2013).
\bibitem{Girvin13}  K. B{\o}rkje, A. Nunnenkamp, J. D. Teufel, and S. M. Girvin, Phys. Rev. Lett. {\bf 111}, 053603 (2013).
\bibitem{Hammerer}  S. P. Tarabrin, H. Kaufer, F. Ya. Khalili, R. Schnabel, K. Hammerer, Phys. Rev. A {\bf 88}, 023809 (2013).
\bibitem{Law} G. -F. Xu and C. K. Law, Phys. Rev. A {\bf 87}, 053849 (2013).

\bibitem{Asjad} M. Asjad,  J. Russ Laser Res.  {\bf 34}, 159 (2013).
\bibitem{Lehnert}T. A. Palomaki,  J. W. Harlow,  J. D. Teufel,  R. W. Simmonds, and  K. W. Lehnert, Nature {\bf 495}, 210 (2013).
\bibitem{Kirchmair} G. Kirchmair, B. Vlastakis, Z. Leghtas, S. E. Nigg, H. Paik, E. Ginossar, M. Mirrahimi, L. Frunzio, S. M. Girvin, and R. J. Schoelkopf, Nature {\bf 495}, 205 (2013).
\bibitem{Ge} W. Ge, M. Al-Amri, H. Nha, M. S. Zubairy, Phys. Rev. A {\bf 88}, 052301 (2013).

\bibitem{Wolf} L. Mandel, E. Wolf, Optical Coherence and Quantum Optics (Cambridge University Press, 1995), Chap. 21.
\bibitem{Wallsbook} D. F. Walls and G. J. Milburn, Quantum Optics (Springer-Verlag, Berlin, 1998), Chap. 8.
\bibitem{GSAbook} G. S. Agarwal, “Quantum Optics”, (Cambridge University Press, 2012), Chap. 3.

\bibitem{spin1}  D. J. Wineland, J. J. Bollinger, W. M. Itano, and D. J. Heinzen, Phys. Rev. A {\bf 50}, 67 (1994).
\bibitem{spin2}  R. McConnell, H. Zhang, S. \'{C}uk, J. Hu, M. H. Schleier-Smith, and V. Vuleti\'{c}, Phys. Rev. A {\bf 88}, 063802 (2013).
\bibitem{sood}  G. A. Garrett, A. G. Rojo, A. K. Sood, J. F. Whitaker and R. Merlin, Science, {\bf 275}, 1638(1997).

\bibitem{BEC0}  L. Duan, M. Lukin, J. I. Cirac, and P. Zoller, Nature {\bf 414}, 413, (2001).
\bibitem{BEC1}  F. Brennecke, S. Ritter, T. Donner, T. Esslinger, Science, {\bf 322}, 235(2008).
\bibitem{BEC2}  J. Lian, L. Yu, J.-Q. Liang, G. Chen, and S. Jia, Scientific Reports {\bf 3}, 3166 (2013).
\bibitem{BEC3}  B. L\"{u}cke, M. Scherer, J. Kruse, L. Pezz\'{e}, F. Deuretzbacher, P. Hyllus, O. Topic, J. Peise, W. Ertmer, J. Arlt, L. Santos, A. Smerzi, and C. Klempt, Science {\bf 334}, 773, (2011).
\bibitem{CavesRMP} C. M. Caves, K. S. Thorne, R. W. P. Drever, V. D. Sandberg, and M. Zimmermann, Rev. Mod. Phys. {\bf 52}, 341 (1980).
\bibitem{Braginsky} V. B. Braginsky, Y. I. Vorontsov, K. S. Thorne, Science {\bf 209}, 547 (1980).
\bibitem{Yurke} B. Yurke, S. L. McCall, and J. R. Klauder, Phys. Rev. A {\bf 33}, 4033 (1986).


\bibitem{sqmirror1}  A. A. Clerk, F. Marquardt, and K. Jacobs, New J. Phys. {\bf 10}, 095010 (2008).
\bibitem{sqmirror2}  K. J\"{a}hne, C. Genes, K. Hammerer, M. Wallquist, E. S. Polzik, and P. Zoller, Phys. Rev. A {\bf 79}, 063819 (2009).
\bibitem{CavesPRL} M. Tsang and C. M. Caves, Phys. Rev. Lett. {\bf 105}, 123601 (2010).
\bibitem{Brooks} D. W. C. Brooks,  T. Botter,  S. Schreppler,  T. P. Purdy,  N. Brahms, and D. M. Stamper-Kurn, Nature {\bf 488}, 476 (2012).
\bibitem{squeezing} A. H. Safavi-Naeini, S. Gr\"{o}blacher, J. T. Hill, J. Chan, M. Aspelmeyer, and O. Painter, Nature (London) {\bf 500}, 185 (2013).
\bibitem{PRX} T. P. Purdy, P.-L. Yu, R. W. Peterson, N. S. Kampel, and C. A. Regal, Phys. Rev. X  {\bf 3}, 031012 (2013).
\bibitem{GSA} M. Asjad, G. S. Agarwal, M. S. Kim, P. Tombesi, G. Di Giuseppe, D. Vitali, arXiv:1309.5485.

\bibitem{Leuch} A. Sizmann and G. Leuch, in Prog. Opt. {\bf 39}, 373 edited by E. Wolf (Elsevier, Amsterdam, 1999).
\bibitem{Vitali} M. Karuza, C. Biancofiore, M. Bawaj, C. Molinelli, M. Galassi, R. Natali, P. Tombesi, G. Di Giuseppe, and D. Vitali, Phys. Rev. A {\bf 88}, 013804 (2013); M. Karuza, C. Molinelli, M. Galassi, C. Biancofiore, R. Natali, P. Tombesi, G. Di Giuseppe, and D. Vitali, New J. Phys. {\bf 14}, 095015 (2012).
\bibitem{app1} C. Jiang, H. Liu, Y. Cui, X. Li, G. Chen, and B. Chen, Opt. Express {\bf 21}, 12165 (2013).
\bibitem{app2} Kenan Qu and G. S. Agarwal, Phys. Rev. A {\bf 87}, 031802 (2013).
\bibitem{app4} J. T. Hill, A. H. Safavi-Naeini, J. Chan, and O. Painter, Nature Commun. {\bf 3}, 1196 (2012); C. Dong, V. Fiore, M. C. Kuzyk, L. Tian, H. Wang, arXiv:1205.2360 (2012); Sh. Barzanjeh, M. Abdi, G. J. Milburn, P. Tombesi, and D. Vitali, Phys. Rev. Lett. {\bf 109}, 130503 (2012).
\bibitem{app5} M. Ludwig, A. H. Safavi-Naeini, O. Painter, and F. Marquardt, Phys. Rev. Lett. {\bf 109}, 063601 (2012); P. K\'{o}m\'{a}r, S. D. Bennett, K. Stannigel, S. J. M. Habraken, P. Rabl, P. Zoller, and M. D. Lukin, Phys. Rev. A {\bf 87}, 013839 (2013).
\bibitem{app6} C. Jiang, H. Liu, Y. Cui, X. Li, G. Chen, and X. Shuai, Phys. Rev. A {\bf 88}, 055801 (2013).
\bibitem{app7} X.-W. Xu and Y.-J. Li, J. Phys. B: At., Mol. Opt. Phys. {\bf 46}, 035502 (2013); H. Flayac and V. Savona, Phys. Rev. A {\bf 88}, 033836 (2013).
\bibitem{app8} M. J. Woolley and A. A. Clerk, Phys. Rev. A {\bf 87}, 063846 (2013).
\bibitem{review} M. Aspelmeyer, T. J. Kippenberg, F. Marquardt, arXiv:1303.0733 (2013).

\bibitem{Hurwitz} A. Hurwitz, On the conditions under which an equation has only roots with negative real parts. Selected Papers on Mathematical Trends in Control Theory. (Dover, New York, 1964).
\bibitem{purity} N. J. Cerf, Gerd Leuchs, and E. S. Polzik, “Quantum Information with Continuous Variables of Atoms and Light”, (World Scientific, 2007), Chap. 1.




\end{thebibliography}
\end{document}